\documentclass[12pt,twoside,leqno]{article}

\usepackage{amsmath}
\usepackage{amsthm}
\usepackage{latexsym}
\usepackage{graphicx}
\usepackage{amssymb}
\usepackage{epstopdf}

\usepackage{epsfig}
\usepackage{booktabs}
\usepackage{pdfpages}

\renewcommand\rightmark{Simulating self-propelled swimmer motion in shear thinning fluids}

\begin{document}
\newtheorem{theorem}{Theorem}
\newtheorem{lemma}{Lemma}
\newtheorem{corollary}{Corollary}

\theoremstyle{remark}
\newtheorem{remark}{Remark}
%\numberwithin{equation}{section}

\parskip 4pt
\abovedisplayskip 7pt
\belowdisplayskip 7pt

\parindent=12pt

\newcommand{\A}{{\bf A}}
\newcommand{\B}{{\bf B}}
\newcommand{\bco}{{\boldsymbol{:}}}
\newcommand{\blambda}{{\boldsymbol{\lambda}}}
\newcommand{\bmu}{{\boldsymbol{\mu}}}
\newcommand{\bn}{{\bf n}}
\newcommand{\bnabla}{{\boldsymbol{\nabla}}}
\newcommand{\bomega}{{\boldsymbol{\omega}}}
\newcommand{\bsigma}{{\boldsymbol{\sigma}}}
\newcommand{\btheta}{{\boldsymbol{\theta}}}
\newcommand{\bg}{{\bf g}}
\newcommand{\bu}{{\bf u}}
\newcommand{\bU}{{\bf U}}
\newcommand{\bv}{{\bf v}}
\newcommand{\bw}{{\bf w}}
\newcommand{\bzero}{{\bf 0}}
\newcommand{\ct}{{\mathcal{T}}}
\newcommand{\cth}{{\mathcal{T}_h}}
\newcommand{\dsum}{{\displaystyle\sum}}
\newcommand{\D}{{\bf D}}
\newcommand{\e}{{\bf e}}
\newcommand{\F}{{\bf F}}
\newcommand{\G}{{\bf G}}
\newcommand{\g}{{\bf g}}
\newcommand{\Gx}{{{\overrightarrow{Gx}}^{\perp}}}
\newcommand{\I}{{\bf I}}
\newcommand{\intbt}{{\displaystyle{\int_{B(t)}}}}
\newcommand{\intG}{{\displaystyle{\int_{\Gamma}}}}
\newcommand{\into}{{\displaystyle{\int_{\Omega}}}}
\newcommand{\intpb}{{\displaystyle{\int_{\partial B}}}}
\newcommand{\lto}{{L^2(\Omega)}}
\newcommand{\no}{{\noindent}}
\newcommand{\obo}{{\Omega \backslash \overline{B(0)}}}
\newcommand{\obt}{{\Omega \backslash \overline{B(t)}}}
\newcommand{\oo}{{\overline{\Omega}}}
\newcommand{\R}{{\text{I\!R}}}
\newcommand{\T}{{\bf T}}
\newcommand{\V}{{\bf V}}
\newcommand{\w}{{\bf w}}
\newcommand{\x}{{\bf x}}
\newcommand{\Y}{{\bf Y}}
\newcommand{\y}{{\bf y}}
\newcommand{\thickhline}{%
	\noalign {\hrule height 2pt}%
}
\newpage
\thispagestyle{empty}

\noindent{\Large\bf On the DLM/FD methods for simulating neutrally buoyant swimmer motion in non-Newtonian shear thinning fluids}
                 
\bigskip
\normalsize \noindent{Ang Li$^a$, Tsorng-Whay Pan$^{b,}$\footnote{Corresponding author: e-mail:  
		pan@math.uh.edu},  and Roland Glowinski$^b$} 
\vskip 2ex
\noindent{\small $^a$ Department of Mathematics, Lane College, Jackson, Tennessee 38301, USA}  
\vskip 1ex
\noindent{\small $^b$ Department of Mathematics, University of Houston, Houston, Texas  77204, USA}  

\vskip 4ex
\noindent {\bf Abstract}
\vskip 1ex 
In this article we discuss the generalization of a Lagrange multiplier based 
fictitious domain (DLM/FD) method to simulating the motion of neutrally buoyant 
particles of non-symmetric shape in  non-Newtonian shear thinning fluids. 
Numerical solutions of steady Poiseuille flow of non-Newtonian shear 
thinning fluids are compared with the exact solutions in a two-dimensional channel. 
Concerning a self-propelled swimmer formed by two disks,
the effect of shear thinning makes the swimmer moving faster and decreases the critical 
Reynolds number (for the moving direction changing to the opposite one) when decreasing 
the value of the power index in the Carreau-Bird model.

\vskip 4ex
\noindent{\it keywords:} Carreau-Bird model; Neutrally buoyant particle;  Self-propelled swimmer;  
Fictitious domain method; Operator splitting; Finite element approximations.

\section{Introduction}
Micro-swimmers, like microorganisms, swimming in a low Reynolds number regime, encounter
stringent constraints due to the dominance of viscous over inertial forces  
\cite{purcell1977, lauga2009}. As a result of kinetic reversibility, Purcell's scallop 
theorem  \cite{purcell1977} rules out reciprocal motion (i.e., strokes with time-reversal 
symmetry) for effective locomotion in the absence of inertia in Newtonian fluids.  
Another important fact is that many biological fluids, including blood, respiratory and 
cervical mucus, have non-Newtonian properties, such as viscoelasticity and shear-thinning 
viscosity (i.e., the viscosity decreases non-linearly with the shear rate \cite{carreau1972} 
and \cite{bird1987}). The physics governing microorganism locomotion at low Reynolds numbers 
in Newtonian fluids is relatively well understood, while low Reynolds propulsion in 
non-Newtonian fluids remains largely unexplored \cite{lauga2009}.
Since the scallop theorem does not hold in non-Newtonian fluids (see, e.g., \cite{lauga2009b} 
and \cite{normand2008}), it should be possible to design and build novel micro-swimmers 
that can move with reciprocal motion in non-Newtonian fluids. A study on a reciprocal sliding 
sphere swimmer in a shear-thinning (non-elastic) fluid \cite{johnson2013} suggests that 
propulsion is achievable by reciprocal motion, in which backward and forward
strokes occur at different rates. Similarly, another recent  study in \cite{qiu2014}
reported that a symmetric micro-scallop, a single-hinge microswimmer, can propel
in shear thickening and shear thinning (non-elastic) fluids at low Reynolds number.
Thus reciprocal swimming mechanism can help in designing biomedical 
microdevices that can propel by a simple actuation scheme in biological  fluids.

In this article,  we have focused on  the effect of shear thinning on the swimmer motion in a
two-dimensional channel. To study such effect by direct numerical simulation, we have adapted 
the Carreau-Bird viscosity model (see \cite{carreau1972} and \cite{bird1987}) for shear thinning 
property. For example, in \cite{huang1997} and \cite{huang1998}, particle settling   in Oldroyd-B 
viscoelastic fluids with shear thinning property was studied numerically by using the Carreau-Bird 
viscosity model. Direct simulation of the motion of particles has been carried out by using an 
arbitrary Lagrangian-Eulerian moving mesh technique with finite element method. The power law, 
which is the other commonly used model, was used to investigate the effects of both 
shear thickening and thinning on the motion of scallop swimmers numerically in  \cite{qiu2014}. 
A finite element method with the dynamic mesh-adaptation for moving boundaries was used to 
study how the scallop swimmer can swim by reciprocal motion at low (not zero) Reynolds number.
For simulating the swimmer motion in non-Newtonian shear thinning fluids, 
we have extended a Lagrange multiplier based fictitious domain (DLM/FD) method developed for
disks in \cite{pan2002jcp} to simulate the motion of a neutrally buoyant non-symmetric swimmer 
in non-Newtonian shear thinning fluids.  This non-symmetric swimmer of two disks with different 
radii is formed by connecting  their mass centers with a massless spring like the one in, e.g., 
\cite{dombrowski2019} and  \cite{dombrowski2020}. To keep the main flavor of the fictitious 
domain methods, i.e.,  the usage of fast solvers to solve the resulting linear systems, we 
have splitted the diffusion term in our proposed methodologies.  Accurate computational 
solutions are obtained  by the proposed methods for Poiseuille flow  of non-Newtonian shear 
thinning fluids in a two-dimensional channel (thanks to the recently published exact 
solutions for such Poiseuille flow by Griffiths in \cite{griffiths2020}).
Then we have further validated our proposed numerical methods by studying the migration of 
56 disks in a two-dimensional channel, which was considered in \cite{huang2000}. 
For the effect of shear thinning on the swimmer motion,
we have obtained that, via direct numerical simulation,   the swimmer 
moves faster and the critical Reynolds number (for  the moving direction changing to the opposite
one) decreases  when decreasing the value of the power index in the Carreau-Bird model. 
The content of this article is as follows: In Section 2 we introduce a
fictitious domain formulation of the model problem
associated with the neutrally buoyant particle cases; then in Section
3 we discuss the time and space discretizations and the related numerical methodology.
We then present and discuss the numerical results in Section 4. Concluding
remarks are given in Section 5.

\section{A fictitious domain formulation of the model problem}

\begin{figure} [ht!]
	\begin{center}
		\leavevmode
		\includegraphics[width=2.5in]{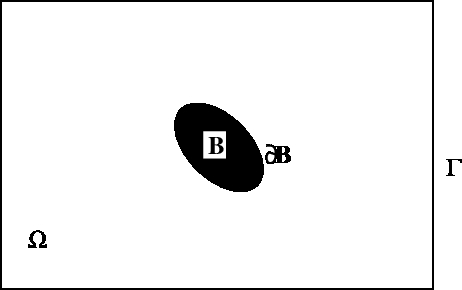}
	\end{center}
	\caption{An example of two-dimensional flow region with one rigid body.}\label{fig.1}
\end{figure}

In \cite{pan2002jcp},  Pan and Glowinski developed a distributed Lagrange multiplier/fictitious 
domain (DLM/FD) method to simulate the motion of neutrally buoyant disks freely moving in a Newtonian 
viscous incompressible fluid in a two dimensional channel. Later they extended in \cite{pan2005crm},
\cite{yang2005jfm}, and \cite{pan2021pof} such DLM/FD method to simulate and investigate the motion of 
neutrally buoyant balls in circular Poiseuille flows. In this article, we have extended the above
DLM/FD  method to study the motion of a self-propelled swimmer in non-Newtonian shear thinning fluids. 
This non-symmetric swimmer of two disks with different radii is formed by connecting  their mass centers
with a massless spring as the one in, e.g., \cite{dombrowski2019} and  \cite{dombrowski2020}.
In this section, we will address  the DLM/FD formulation for such problem.  
Let $\Omega \subset \R^2$ be a rectangular region. We suppose that $\Omega$ is filled with a
{\it non-Newtonian viscous incompressible} fluid of {\it density} $\rho_f$  and contains
a moving neutrally buoyant rigid particle  $B$ centered at $\G=\{G_1, G_2\}^T$
of {\it density} $\rho_f$ as shown in Figure \ref{fig.1}; the flow is modeled
by the {\it Navier-Stokes equations} and the motion of $B$
is described by the {\it Euler-Newton's equations}. Following the DLM/FD formulation 
developed in \cite{pan2002jcp},  we define
\begin{eqnarray*}
&&W_{0, p} = \{\bv|\bv \in (H^1(\Omega))^2, \ \bv = {\bf 0} \
\text{\it on the top and bottom boundary of $\Omega$ and} \\
&&\hskip 55pt \bv \ \text{\it is periodic in the $x_1$ direction} \},\\
&&L_{0}^2  = \{q|q \in L^2(\Omega), \int_{\Omega} q\, d\x=0,\},\\
&&\Lambda_0(t) = \{\bmu| \bmu \in (H^1(B(t)))^2, <\bmu,{\bf e}_i>_{B(t)}=0, \ i=1, 2,
<\bmu, \Gx >_{B(t)} = 0\}
\end{eqnarray*}
with  ${\bf e}_1=\{1, 0\}^T$, ${\bf e}_2=\{0, 1\}^T$, $\Gx =\{-(x_2-G_2),x_1-G_1\}^T$ and
$<\cdot,\cdot>_{B(t)}$ an inner product on $\Lambda_0(t)$ which can
be the standard inner product on $(H^1(B(t)))^2$ (see \cite{glowinski2001}, Section 5, for further
information on the choice of $<\cdot,\cdot>_{B(t)}$). Then the fictitious domain formulation
with distributed Lagrange multipliers for flow around a freely moving  neutrally buoyant
particle  is as follows
\begin{eqnarray*}
&&For \ a.e. \ t>0, \ find \
\bu(t) \in W_{0, p}, \ p(t) \in L_{0}^2, \ \V_{\G}(t) \in
\R^2, \ \G(t) \in \R^2, \\
&&\omega(t) \in \R, \ \blambda(t) \in \Lambda_0(t) \ \ \text{\it such that}
\end{eqnarray*}
\begin{eqnarray}
&&\begin{cases}
\rho_f \into \left[\dfrac{\partial \bu}{\partial t}+(\bu \cdot \bnabla) \bu \right] \cdot \bv\ d\x
+ 2 \into \eta \ \D(\bu) \boldsymbol{:} \D(\bv)\ d\x
- \into p \bnabla \cdot \bv\, d\x \\
\ \ \ -<\blambda, \bv>_{B(t)} = \rho_f \into \g \cdot \bv\, d\x + \into \F \cdot \bv\, d\x, \
\forall \bv \in W_{0, p},\end{cases} \label{eqn:2.1}
\end{eqnarray}
\begin{eqnarray}
&& \into q \bnabla \cdot \bu(t) d\x = 0, \ \forall q \in L^2(\Omega),\label{eqn:2.2} \\
&&<\bmu, \bu(t) >_{B(t)} =<\bmu, \bu_p(t)>_{B(t)}, \ \forall \bmu \in \Lambda_0(t),\label{eqn:2.3}\\
&&\dfrac{d\G}{dt}=\V_{\G},  \label{eqn:2.4} \\
&& \V_{\G}(0) = \V_{\G}^0, \ \omega(0) = \omega^0, \ \G(0) = \G^0=\{G^0_1, G^0_2\}^T,
\label{eqn:2.5}\\
&&\bu(\x, 0) = {\overline \bu}_0(\x) = \begin{cases}
\bu_0(\x), \ \forall \x \in \obo, \\
\V_{\G}^0 + \omega^0 \{-(x_2-G^0_2),x_1-G^0_1\}^T, \ \forall \x \in \overline{B(0)},
\end{cases}
\label{eqn:2.6}
\end{eqnarray}

\noindent where
$\bu$ and $p$ denote velocity and pressure, respectively, $\eta(\dot{\gamma})$ is the 
fluid viscosity,
$\blambda$ is a Lagrange multiplier, $\D(\bv) = (\bnabla \bv + \bnabla \bv^T)/2$,
$\g$ is gravity, $\F$ is the pressure gradient pointing in the $x_1$ direction for
particle moving in a Poiseuille flow,
$\V_{\G}$ is the {\it translation velocity} of the particle $B$, and
$\omega$ is the {\it angular velocity} of $B$.
We suppose that the {\it no-slip} condition holds on $\partial B$.
We also use, if necessary, the notation $\phi(t)$ for the function $\x \to
\phi(\x,t)$.  {In equation (\ref{eqn:2.3}), $\bu_p(t)$ is a part of actual particle motion 
given as follows
\begin{equation*}
\bu(\x,t)=\V_{\G}(t)+\omega(t) \{-(x_2-G_2),x_1-G_1\}^T+\bu_p(\x,t).
\end{equation*}
Thus we have $\bu_p={\bf 0}$ if $B$ is a single piece neutrally buoyant rigid 
particle like a disk suspended in fluid; but for $B$ as a swimmer formed by two neutrally buoyant disks
considered in this article, $\bu_p(t)$ is a given 
reciprocal motion  of the two disks with respect to the swimmer mass center (see Section 4).}
Shear thinning can be easily added to the above model by using
the Carreau-Bird viscosity model (see, \cite{carreau1972} and \cite{bird1987})
\begin{equation}
\dfrac{\eta-\eta_{\infty}}{\eta_0-\eta_{\infty}}=(1+(\lambda_3\dot{\gamma})^2)^{(n-1)/2}, \label{eqn:2.6a}
\end{equation}
where $\eta_0$ (resp., $\eta_{\infty}$) is the fluid viscosity at zero shear rate (resp.,  
infinite shear rate), $\lambda_3$ is the relaxation time,  $\dot{\gamma}$ is determined by the second 
invariant of the rate of strain tensor $(\dot{\gamma})^2=(\D(\bu):\D(\bu))/2$, and the power index 
$n$ is in $(0,1]$ (if $n=1$, the fluid is a Newtonian fluid with constant viscosity $\eta_0$).

%
%\begin{remark}
%The hydrodynamical forces and torque imposed on the rigid body
%by the fluid are built in (\ref{eqn:2.1})-(\ref{eqn:2.6})
%implicitly (see \cite{012, 013} for details), thus
%we do not need to compute them explicitly in the simulation.
%Since in (\ref{eqn:2.1})-(\ref{eqn:2.6}) the flow field is
%defined on the entire domain $\Omega$, it can be computed
%with a simple structured grid.
%\end{remark}
\begin{remark}
In (\ref{eqn:2.3}), the rigid body motion in the region occupied
by the particle is enforced via Lagrange multipliers $\blambda$.
To recover the translation velocity $\V_{\G}(t)$  and the angular
velocity $\omega(t)$, we solve the following equations
\begin{equation}
\begin{cases}
<{\bf e}_i, \bu(t)-\V_{\G}(t)- \omega(t) \ \Gx >_{B(t)}=0,\ for \ i=1, 2,\\
<\Gx , \bu(t)-\V_{\G}(t)- \omega(t) \ \Gx >_{B(t)}=0.
\end{cases}\label{eqn:2.7}
\end{equation}
\end{remark}

%\begin{remark}
%In (\ref{eqn:2.1}), $2 \into \D(\bu) : \D (\bv)\, d\x$ can be replaced
%by $\into \bnabla \bu : \bnabla \bv\, d\x$ since $\bu$ is divergence free and
%in $W_{0,p}$. Also the gravity $\g$
%in (\ref{eqn:2.1}) can be absorbed into the pressure term.
%\end{remark}

\section{Space approximation and time discretization}

Concerning the finite element based  space
approximation of problem (\ref{eqn:2.1})-(\ref{eqn:2.6}), we use $P_1$-$iso$-$P_2$ 
and $P_1$ finite elements for the velocity field and pressure, respectively 
(see, e.g., \cite{bristeau1987} and \cite{glowinski2003}(Chapter 5)).
More precisely with $h$ a space discretization step, we introduce a finite
element triangulation $\cth$ of $\oo$ and then $\ct_{2h}$ a
triangulation twice coarser. Then we approximate $W_{0, p}$, $L^2$ and
$L^2_{0}$ by the following finite dimensional spaces
\begin{eqnarray}
&&W_{0,h}=\{\bv_h | \bv_h \in (C^0(\overline{\Omega}))^2, \
\bv_h|_T \in P_1\times P_1, \ \forall T\in \ct_h,  \ \bv_h = {\bf 0} \
\text{\it on the}\label{eqn:3.1}\\
&&\hskip 60pt \text{\it top and bottom boundary of $\Omega$ and} \ \bv \ \text{\it is periodic in the $x_1$
direction} \ \}, \nonumber
\end{eqnarray}
\begin{eqnarray}
L^2_{h}=\{q_h &|& q_h\in C^0(\overline{\Omega}), \ q_h|_T\in P_1,
\ \forall T\in \ct_{2h}, \ q_h \ \text{\it is periodic}\label{eqn:3.2} \\
&&\text{\it in the $x_1$ direction}\}, \nonumber
\end{eqnarray}
and
\begin{equation}
L^2_{0,h}=\{q_h|q_h\in L^2_{h}, \ \int_{\Omega} q_h \, d\x=0 \},
\label{eqn:3.3}
\end{equation}
respectively; in (\ref{eqn:3.1})-(\ref{eqn:3.3}), $P_1$ is the space of
polynomials in two variables of degree $\le 1$.

\begin{figure}[th!]
\begin{center}
\leavevmode
	\includegraphics[width=2.5in]{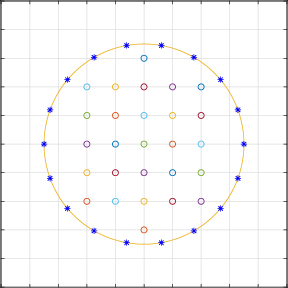}
\end{center}
\caption{An example of set of collocation points chosen inside the disk and at its boundary for enforcing
the rigid body motion.}\label{fig.2}	
\end{figure}

A finite dimensional space approximating $\Lambda_0(t)$
is defined in the following:  let $\{\x_i\}_{i=1}^{N}$ be a set of points
covering $\overline{B(t)}$; we define then
\begin{equation}
\Lambda_{h}(t)=\{\bmu_h|\bmu_h=\dsum_{i=1}^{N}\bmu_i
\delta(\x-\x_i), \ \bmu_i\in\R^2, \ \forall i=1,...,N\},
\label{eqn:3.4}
\end{equation}
where $\delta(\cdot)$ is the Dirac measure at $\x={\bf 0}.$
Then, instead of the inner product of $(H^1(B(t)))^2$
we shall use $<\cdot,\cdot>_{B_h(t)}$ defined by
\begin{equation}
<\bmu_h,\bv_h>_{B_h(t)} = \dsum_{i=1}^{N}
\bmu_i\cdot\bv_h(\x_i), \ \forall \bmu_h \in \Lambda_{h}(t),
\ \bv_h\in W_{0,h}.
\label{eqn:3.5}
\end{equation}
Then we approximate $\Lambda_0(t)$ by
\begin{equation}
\Lambda_{0,h}(t) =\{\mu_h| \mu_h \in \Lambda_h(t), \ <\mu_h,{\bf e}_i>_{B_h(t)}=0,\
i=1, 2, \ <\mu_h, \Gx >_{B_h(t)}=0 \}.\label{eqn:3.4a}
\end{equation}
A typical choice of points for defining (\ref{eqn:3.4}) is to take the grid points of the velocity 
mesh internal to the particle $B$ and whose distance to the boundary of $B$ is greater than, e.g. $h$, 
and to complete with selected points from the boundary of $B(t)$ (see Figure \ref{fig.2}).

Using the above finite dimensional spaces leads to the
following approximation of problem   (\ref{eqn:2.1})-(\ref{eqn:2.6}):
\begin{eqnarray*}
&&For \ a.e. \ t>0, \ find \
\bu(t) \in W_{0,h}, \ p(t) \in L_{0,h}^2, \ \V_{\G}(t) \in
\R^2, \ \G(t) \in \R^2, \\
&&\omega(t) \in \R, \ \blambda_h(t) \in \Lambda_{0,h}(t) \ \ \text{\it such that}
\end{eqnarray*}
\begin{eqnarray}
&&\begin{cases}
\rho_f \into \left[\dfrac{\partial \bu_h}{\partial t}  +
  (\bu_h \cdot \bnabla) \bu_h \right] \cdot \bv\, d\x
+  2 \into \eta \ \D(\bu_h) \boldsymbol{:} \D(\bv)\ d\x    \\
\hskip 20pt - \into p_h \bnabla \cdot \bv\, d\x - <\blambda_h, \bv>_{B_h(t)}
=  \into \F \cdot \bv\, d\x, \
\forall \bv \in W_{0,h},\end{cases} \label{eqn:3.6}\\
&& \into q \bnabla \cdot \bu_h(t) d\x = 0, \ \forall q \in L^2_{h},\label{eqn:3.7} \\
&&<\bmu, \bu_h(t) >_{B_h(t)} =<\bmu, \bu_{p}(t) >_{B_h(t)} , \ \forall \bmu \in \Lambda_{0,h}(t),\label{eqn:3.8}\\
&&\dfrac{d\G}{dt}=\V_{\G},  \label{eqn:3.9} \\
&& \V_{\G}(0) = \V_{\G}^0, \ \omega(0) = \omega^0, \ \G(0) = \G^0=\{G^0_1, G^0_2\}^T,
\label{eqn:3.10}\\
&&\bu_h(\x, 0) ={\overline\bu}_{0,h}(\x) \label{eqn:3.11} \
(\text{with} \bnabla \cdot {\overline\bu}_{0,h} =0).
\end{eqnarray}

Applying a first order operator splitting scheme, namely the Lie scheme \cite{chorin1978} (see also \cite{glowinski2003} and \cite{ogy2016} (Chapter 2)),  
to discretize equations (\ref{eqn:3.6})-(\ref{eqn:3.11}) in time, we obtain (after
dropping some of the subscripts $h$):
\vskip 1ex
\begin{equation}
\bu^0={\overline\bu}_{0,h}, \ \V_{\G}^0, \ \omega^0, \ and \ \G^0 \ are \ given;
\label{eqn:3.12}
\end{equation}
%\vskip 1ex
\noindent{\it For $n \ge 0$, knowing $\bu^n$, $\V_{\G}^n$, $\omega^n$ and $\G^n$,
compute  $\bu^{n+1/6}$ and $p^{n+1/6}$ via the solution of}
\begin{equation}
\begin{cases}
\displaystyle \rho_f \into
\frac{\bu^{n+1/6} - \bu^n}{\triangle t} \cdot \bv\, d\x
- \into p^{n+1/6} \bnabla \cdot \bv\, d\x=0,
\  \forall \bv \in W_{0,h}, \\
\displaystyle \into q \bnabla \cdot \bu^{n+1/6}\,d\x=0,
\ \forall q \in L^2_{h}; \ \bu^{n+1/6} \in W_{0,h},
\  p^{n+1/6} \in L^2_{0,h}.
\end{cases}  \label{eqn:3.13}
\end{equation}
\vskip 2ex
\noindent {\it Then compute $\bu^{n+2/6}$ via the solution of}
\vskip 1ex
\begin{eqnarray}
&&\begin{cases}
\displaystyle  \into \frac{\partial \bu}{\partial t} \cdot \bv\, d\x
 + \into ( \bu^{n +1/6} \cdot \bnabla ) \bu \cdot \bv\, d\x =0, \
\forall \bv \in W_{0,h}, \ on \ (t^n, t^{n+1}), \\
\displaystyle  \ \bu(t^n)= \bu^{n+1/6}; \ \ \bu(t) \in W_{0,h},
\end{cases} \label{eqn:3.14} \\
&&\bu^{n+2/6} = \bu(t^{n+1}). \label{eqn:3.15}
\end{eqnarray}
\vskip 4ex
\noindent {\it Next, compute $\bu^{n+3/6}$ via the solution of}
\begin{equation}
\begin{cases}
\displaystyle\rho_f \into\frac{\bu^{n+3/6} - \bu^{n+2/6}}{\triangle t} \cdot \bv\, d\x
+  2 \alpha \into \eta(\dot{\bu^{n+2/6}}) \ \D(\bu^{n+3/6}) \boldsymbol{:} \D(\bv)\ d\x  \\ 
\hskip 20pt = \into \F \cdot \bv\, d\x,  
\forall \bv \in W_{0,h}; \  \ \bu^{n+3/6} \in W_{0,h}.
\end{cases}  \label{eqn:3.16}
\end{equation}
\vskip 1ex
\noindent {\it Now predict the position and the translation
velocity of the center of mass of the particles as follows:}
\vskip 1ex
{\it Take $\V_{\G}^{n+\frac{4}{6},0}=\V_{\G}^{n}$ and $\G^{n+\frac{4}{6},0}=\G^{n}$;
then predict the  new position of the particle
via the following sub-cycling and predicting-correcting technique:}
\vskip 1ex
\begin{eqnarray}
&&\kern -10pt \text{\it For $k=1, \dots, N$, compute}\, \nonumber\\
&&{\widehat \V_{\G}}^{n+\frac{4}{6},k} = \V_{\G}^{n+\frac{4}{6},k-1}
+\F^r(\G^{n+\frac{4}{6},k-1}) {\triangle t}/{2N},\label{eqn:3.17}\\
&& {\widehat \G}^{n+\frac{4}{6},k} = \G^{n+\frac{4}{6},k-1} +
  ({\widehat \V_{\G}}^{n+\frac{4}{6},k} + \V_{\G}^{n+\frac{4}{6},k-1}){\triangle t}/{4N},
   \label{eqn:3.18}\\
&& \V_{\G}^{n+\frac{4}{6},k} = \V_{\G}^{n+\frac{4}{6},k-1}
 + (\F^r({\widehat \G}^{n+\frac{4}{6},k})+\F^r(\G^{n+\frac{4}{6},k-1})){\triangle t}/{4N},
   \label{eqn:3.19}\\
&& \G^{n+\frac{4}{6},k} = \G^{n+\frac{4}{6},k-1}
+ ( \V_{\G}^{n+\frac{4}{6},k}+\V_{\G}^{n+\frac{4}{6},k-1} ){\triangle t }/{4N},
  \label{eqn:3.20}\\
&&\kern -10pt  \text{\it enddo;} \nonumber\\
&&\kern -10pt  \text{\it and let \ $\V_{\G}^{n+\frac{4}{6}}=\V_{\G}^{n+\frac{4}{6},N},
\ \G^{n+\frac{4}{6}}=\G^{n+\frac{4}{6},N}$.}\, \label{eqn:3.21}
\end{eqnarray}
\vskip 1ex
\noindent {\it Now, compute $\bu^{n+5/6}$, ${\blambda}^{n+5/6}$,
$\V_{\G}^{n+5/6}$,  and $\omega^{n+5/6}$ via the solution of}
\vskip 1ex
\begin{equation}
\begin{cases}
\displaystyle\rho_f  \into   \frac{\bu^{n+5/6}
- \bu^{n+3/6}}{\triangle t}  \cdot \bv\, d\x
+2 \beta   \into \eta(\dot{\bu^{n+3/6}}) \ \D(\bu^{n+5/6}) \boldsymbol{:} \D(\bv)\ d\x   \\
\ =<\blambda, \bv >_{B_h^{n+4/6}},
\ \forall \bv \in W_{0,h},  \\
 <\bmu, \bu^{n+5/6} >_{B_h^{n+4/6}}= <\bmu, \bu_{p}(t^{n+1}) >_{B_h^{n+4/6}},
\ \forall {\bmu} \in \Lambda_{0,h}^{n+4/6};
\displaystyle  \ \bu^{n+5/6} \in W_{0,h},
{\blambda}^{n+5/6} \in  \Lambda_{0, h}^{n+4/6},
\end{cases}  \label{eqn:3.22}
\end{equation}
\vskip 1ex
\noindent{\it and solve for $\V_{\G}^{n+5/6}$ and $\omega^{n+5/6}$ from}
\begin{equation}
\begin{cases}
<{\bf e}_i, \bu^{n+5/6} -\V_{\G}^{n+5/6} -
\omega^{n+5/6} \ {\overrightarrow{G^{n+4/6}x}}^{\perp} >_{B_h^{n+4/6}}=0,
\ for \ i=1, 2,\\
<{\overrightarrow{G^{n+4/6}x}}^{\perp} , \bu^{n+5/6} -\V_{\G}^{n+5/6} -
\omega^{n+5/6} \ {\overrightarrow{G^{n+4/6}x}}^{\perp} >_{B_h^{n+4/6}}=0,
\end{cases}
\label{eqn:3.23}
\end{equation}
\vskip 1ex
\noindent {\it Finally, take $\V_{\G}^{n+1,0}=\V_{\G}^{n+5/6}$ and $\G^{n+1,0}=\G^{n+4/6}$;
then predict the final position and translation velocity
as follows:}
\vskip 1ex
\begin{eqnarray}
&&\kern -10pt \text{\it For $k=1, \dots, N$, compute}\, \nonumber\\
&& {\widehat \V_{\G}}^{n+1,k} = \V_{\G}^{n+1,k-1} + \F^r(\G^{n+1,k-1}){\triangle t}/{2N},
    \label{eqn:3.24}\\
&&{\widehat \G}^{n+1,k} = \G^{n+1,k-1} +
({\widehat \V_{\G}}^{n+1,k} + \V_{\G}^{n+1,k-1}){\triangle t}/{4N},\label{eqn:3.25}\\
&& \V_{\G}^{n+1,k} = \V_{\G}^{n+1,k-1}
 +(\F^r({\widehat \G}^{n+1,k})+\F^r(\G^{n+1,k-1})){\triangle t}/{4N}, \label{eqn:3.26}\\
&& \G^{n+1,k} = \G^{n+1,k-1} + ( \V_{\G}^{n+1,k}+ \V_{\G}^{n+1,k-1} ){\triangle t }/{4N},
\label{eqn:3.27}\\
&& \kern -10pt \text{\it enddo;} \nonumber
\end{eqnarray}
{\it and let \ $\V_{\G}^{n+1}=\V_{\G}^{n+1,N}, \ \G^{n+1}=\G^{n+1,N}$; and set
$\bu^{n+1}=\bu^{n+5/6}$,  $\omega^{n+1}=\omega^{n+5/6}$.}
\vskip 2ex

In algorithm (\ref{eqn:3.12})-(\ref{eqn:3.27}),  we have
$t^{n+s}=(n+s)\triangle t$,  $\Lambda_{0,h}^{n+s}=\Lambda_{0,h}(t^{n+s})$,
$B_h^{n+s}$ is the region occupied by the particle centered at $\G^{n+s}$, and
$\F^r$ is a short range repulsion force which prevents the particle/particle and
particle/wall penetration (see, e.g., \cite{012, glowinski2001}).
Finally, $\alpha$ and $\beta$ verify $\alpha+\beta=1$; we have chosen
$\alpha=1$ and $\beta=0$ in the numerical simulations discussed later.

\subsection{Solutions of the subproblems (\ref{eqn:3.13}), (\ref{eqn:3.14}),
(\ref{eqn:3.16}) and (\ref{eqn:3.22})}

The degenerated quasi-Stokes problem (\ref{eqn:3.13}) is solved by a
preconditioned conjugate gradient method introduced in \cite{011},
in which the discrete elliptic problems from the preconditioning are solved by
a  matrix-free fast solver from FISHPAK by Adams et al. in \cite{001}.
The advection problem (\ref{eqn:3.14}) for the velocity field  is solved by
a wave-like equation method as in \cite{007, 022}.
To enforce the rigid body motion inside the region occupied by the particles,
we employ the conjugate gradient method discussed in  \cite{pan2002jcp} for problem (\ref{eqn:3.22}).
Problem (\ref{eqn:3.16}) is a classical discrete elliptic problem. To solve
problem  (\ref{eqn:3.16}) by the same matrix-free fast solver, we have modified the equation
as follows:
\begin{equation}
\begin{cases}
\displaystyle\rho_f \into\frac{\bu^{n+3/6} - \bu^{n+2/6}}{\triangle t} \cdot \bv\, d\x
+ \alpha \eta_0 \into \bnabla\bu^{n+3/6} {\bf :} \bnabla\bv\,d\x   
=  \into \F \cdot \bv\, d\x\\
\hskip 20pt + \alpha  \eta_0 \into \bnabla\bu^{n+2/6} {\bf :} \bnabla\bv\,d\x -  2 \alpha \into \eta(\dot{\bu^{n+2/6}}) \ \D(\bu^{n+2/6}) \boldsymbol{:} \D(\bv)\ d\x,  \\
\forall \bv \in W_{0,h}; \  \ \bu^{n+3/6} \in W_{0,h}.
\end{cases}  \label{eqn:3.16a}
\end{equation}
In problem (\ref{eqn:3.16a}), the value of $\bnabla\bu^{n+2/6}$ at each mesh node is obtained via a 
super-convergent scheme discussed in \cite{whiteman1991}.  In this article, all computational results 
of shear thinning fluids were obtained  by algorithm (\ref{eqn:3.12})-(\ref{eqn:3.27}) with the 
above subproblem  (\ref{eqn:3.16a}) instead of problem (\ref{eqn:3.16}).

\section{Numerical experiments and discussion}

\subsection{Poiseuille flow of non-Newtonian shear thinning fluids}
In the first validation case, we have considered the pressure driven Poiseuille flow of a non-Newtonian 
shear thinning fluid. The computational results are validated by comparing them with the  exact 
solutions recently reported in \cite{griffiths2020}. Since there is no particle involved in the 
computation, the computational results are obtained from  steps (\ref{eqn:3.12}), (\ref{eqn:3.13}), 
(\ref{eqn:3.14}), (\ref{eqn:3.15}), and (\ref{eqn:3.16a}) in algorithm 
(\ref{eqn:3.12})-(\ref{eqn:3.27}).

The computational domain is $\Omega=(0,L)\times(-H,H)=(0, 8)$ $\times$ $(-1, 1)$. 
We have used structured triangular meshes in all computations. The mesh sizes for 
the velocity field are $h_v=1/32$, 1/64, and 1/96, while the mesh size for the pressure is 
$h_p=2 h_v$. The time step is chosen to be $\triangle t=0.001$. The  pressure gradient 
pointing in the $x_1$ direction is $(F_1, F_2)^T=(-G, 0)^T=(-1,0)^T$  so that the flow
moves from  left to right. The flow velocity is periodic in the $x_1$ (horizontal) direction
and zero at the top and bottom boundary of the domain. The initial flow velocity is ${\bf 0}$.  
The fluid density is 1 and the fluid viscosities  are $\eta_0=1$ and $\eta_{\infty}=0$.
The three values of the power index $n$ considered in this case are 1/3, 1/2, and 2/3. 
Then as given in \cite{griffiths2020}, for the power index $n=1/3$ in  (\ref{eqn:2.6a}), the exact solution 
$(u_1,0)$ and the associated relaxation time $\lambda_3$ of the above shear thinning Poiseuille 
flow problem are given by
\begin{eqnarray*}
&&\lambda_3=\sqrt[6]{2}\sqrt{3}\eta_0/GH=\sqrt[6]{2}\sqrt{3},\\
&&u_1(Y)=\dfrac{\sqrt[3]{2}GH^2}{4 \eta_0}
\left[ \dfrac{3}{\sqrt[3]{4}} + 1-Y^4 \right. \\
&& \hskip 60pt \left. + \dfrac{ (3-\sqrt{1+8Y^6}) (1+\sqrt{1+8Y^6})^{2/3}-2 Y^2 (3+\sqrt{1+8Y^6})}{4 (1+\sqrt{1+8Y^6})^{1/3}} \right],
\end{eqnarray*}
where $Y=y/H$ and $-H\le y\le H$. For $n=1/2$, they are
\begin{eqnarray*}
	&&\lambda_3=\sqrt{2}\eta_0/(\sqrt[4]{3}GH)=\sqrt{2}/\sqrt[4]{3},\\
	&&u_1(Y)= \dfrac{ GH^2 (\sqrt{3+Y^4}-2Y^2) (Y^2+\sqrt{3+Y^4})^{1/2}} {3 \sqrt[4]{3} \eta_0},
\end{eqnarray*}
and for $n=2/3$, they are
\begin{eqnarray*}
	&&\lambda_3=\sqrt{3}\eta_0/(\sqrt[3]{2}GH)=\sqrt{3}/\sqrt[3]{2},\\
	&&u_1(Y)=\dfrac{GH^2}{5\sqrt[6]{2} \eta_0}
	\left[  \sqrt{2}\right. \\
&&\hskip 10pt \left. + 
  \dfrac{ [(1+\sqrt{1-Y^6})^{4/3}+Y^4-3Y^2(1+\sqrt{1-Y^6})^{2/3}][ (1+\sqrt{1-Y^6})^{2/3} + Y^2]^{1/2} } { (1+\sqrt{1-Y^6})^{5/6}} \right],
\end{eqnarray*}
\vskip 2ex
\begin{center}
\hskip 20pt \vbox{\tabskip=0pt \offinterlineskip
\def\tablerule{\noalign{\hrule}}
\halign to 425.4pt{\vrule height 14pt depth 10pt width0pt#& \vrule#\tabskip=0.35em plus0em&
#& \vrule#&  #& \vrule#&  #& \vrule#&  #& \vrule#&  #& \vrule#&  #& \vrule#&   #& \vrule#& #& \vrule#\tabskip=0pt\cr\tablerule
%each {\hfil#& \vrule#&} align one vertical line
&& \hfil $n$ \hfil && \hfil $h$ \hfil &&\omit\hidewidth $L^2$-error  \hidewidth &&\omit\hidewidth order \hidewidth
&& \hfil $n$ \hfil && \hfil $h$ \hfil &&\omit\hidewidth $L^2$-error  \hidewidth &&\omit\hidewidth order \hidewidth
&\cr\tablerule
&& 1/3 && 1/32 &&8.510003368$\times 10^{-3}$ &&        && 1/2 && 1/32 &&1.405723487$\times 10^{-4}$ &&  & \cr\tablerule
&& 1/3 && 1/64 &&2.475437454$\times 10^{-3}$ && 1.833  && 1/2 && 1/64 &&3.667227424$\times 10^{-5}$ && 1.938 & \cr\tablerule
&& 1/3 && 1/96 &&1.157654939$\times 10^{-3}$ && 1.815  && 1/2 && 1/96 &&1.633156504$\times 10^{-5}$ && 1.959 & \cr\tablerule
&& 2/3 && 1/32 &&7.847740427$\times 10^{-5}$ &&       && && &&  && & \cr\tablerule
&& 2/3 && 1/64 &&1.990780506$\times 10^{-5}$ && 2.008 && && &&  && & \cr\tablerule
&& 2/3 && 1/96 &&8.704140007$\times 10^{-6}$ && 2.001 && && &&  && & \cr\tablerule
}}

\vskip 2ex
\begin{minipage}{6in}
Table 1. $L^2$-errors for the steady state velocity field  of Poiseuille flow  for the power indices $n=1/3$, 1/2, and 2/3.
\end{minipage}
\end{center}

\begin{figure}[th!]
	\begin{center}
		\leavevmode
		\includegraphics[width=5.25in]{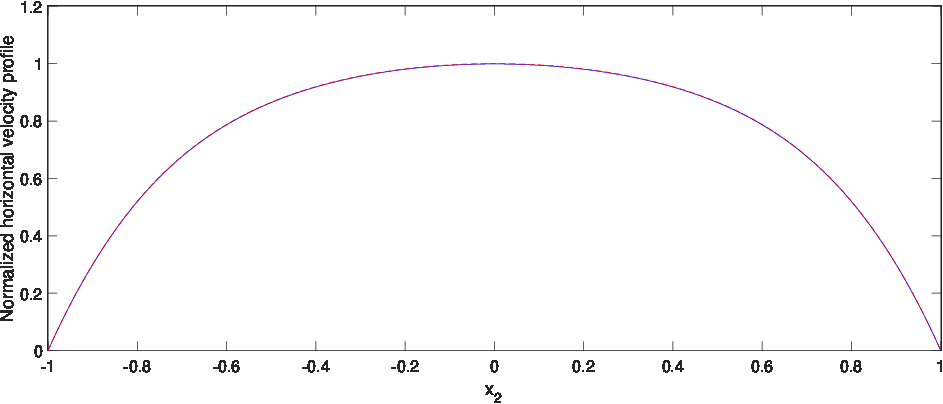}
	\end{center}
	\caption{Normalized horizontal velocity profiles obtained for $h=1/96$ (red dashed line) and  exact 
		solution (blue solid line)  for $n=1/3$.}\label{fig.3}	
\end{figure}
\begin{figure}[th!]
	\begin{center}
		\leavevmode
		\includegraphics[width=5.25in]{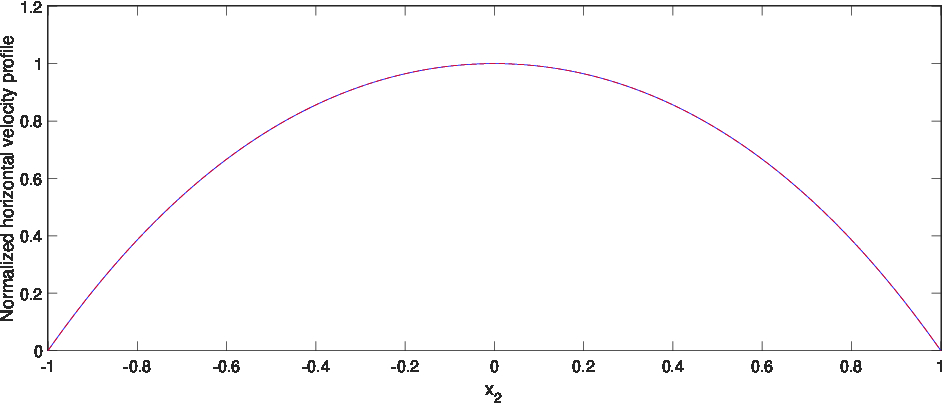}
	\end{center}
	\caption{Normalized horizontal velocity profiles obtained for $h=1/96$ (red dashed line) and  exact 
		solution (blue solid line)  for $n=1/2$.}\label{fig.4}	
\end{figure}
\begin{figure}[th!]
\begin{center}
\leavevmode
\includegraphics[width=5.25in]{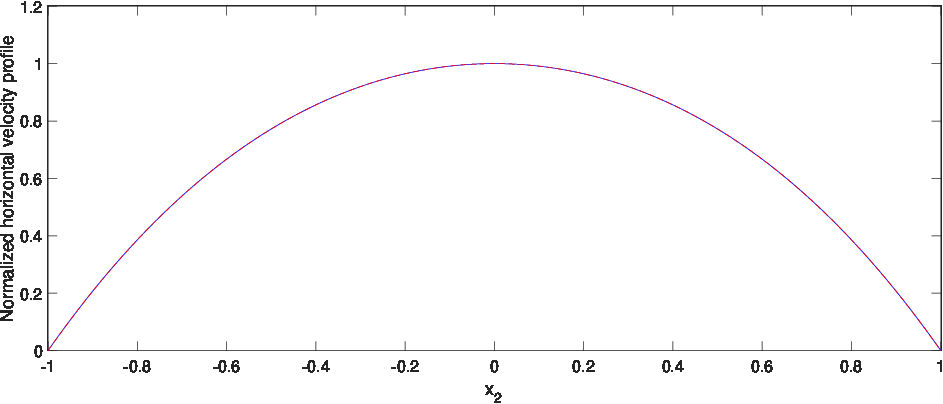}
\end{center}
\caption{Normalized horizontal velocity profiles obtained for $h=1/96$ (red dashed line) and exact
 solution (blue solid line)  for $n=2/3$.}\label{fig.5}	
\end{figure}
The  numerical results obtained with $h=1/96$ and the actual profiles are shown in Figures \ref{fig.3}, 
\ref{fig.4}, and  \ref{fig.5} for the steady state horizontal profile of Poiseuille flow. 
The $L^2$-errors of steady numerical solutions in Table 1 show that the expected error order for 
a $P_1$ finite element approximation has been obtained.

\begin{figure}[th!]
\begin{center}
\leavevmode
(a) \includegraphics[width=5.1in]{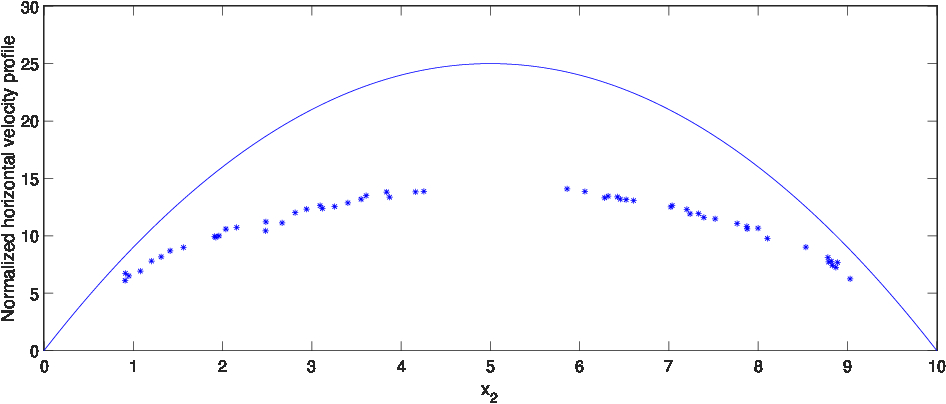}\\
(b) \phantom{1} \includegraphics[width=4.7in]{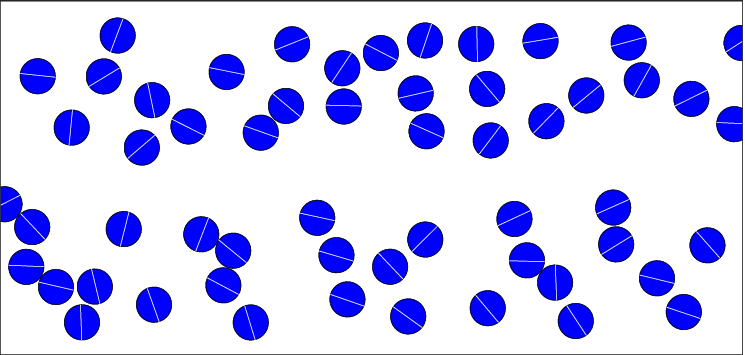}
\end{center}
\caption{Migration of 56 neutrally buoyant identical particles in the pressure driven flow
of a Newtonian fluid without 
shear thinning ($n=1.0$). (a) Velocity profile of the fluid without particles (top curve) and 
velocities of particles (bottom curve); (b) particle positions in the channel  at time $t=100$. }\label{fig.6}	
\end{figure}

\begin{figure}[!tp]
\begin{center}
\leavevmode
\includegraphics[width=5.4in]{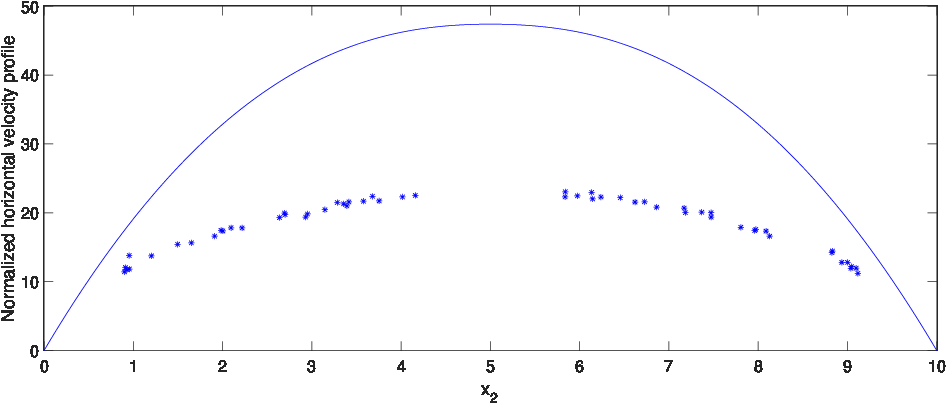}\\
\includegraphics[width=5.4in]{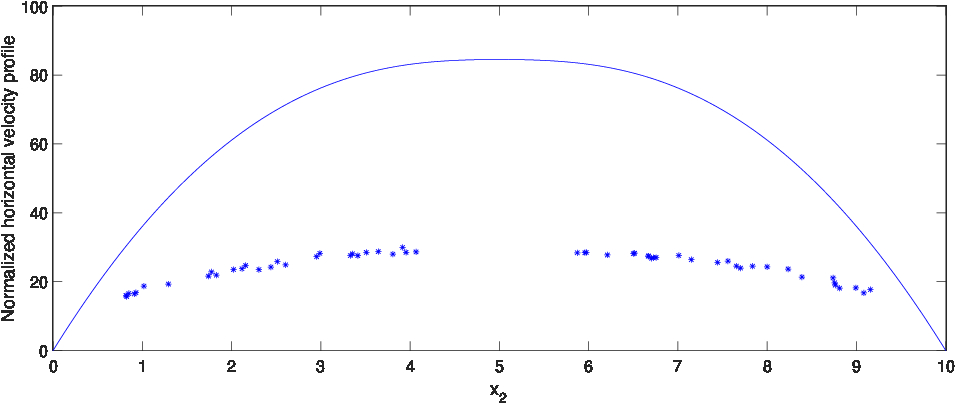}\\
\includegraphics[width=5.4in]{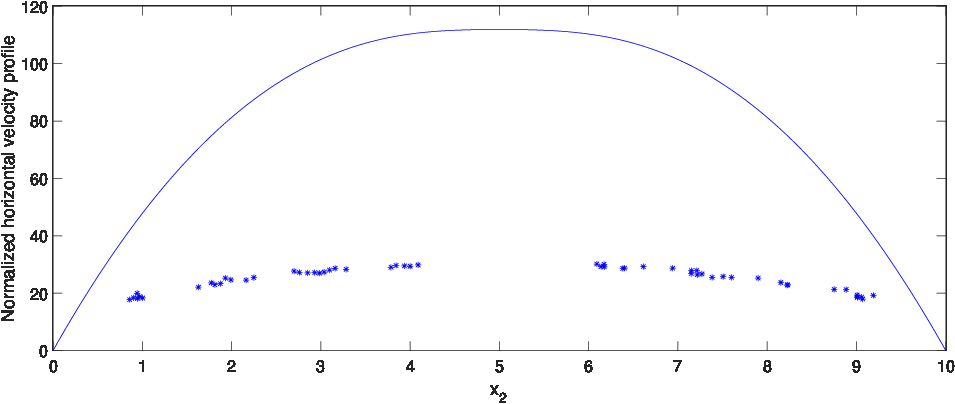}	
\end{center}
\caption{Velocity profile of the fluid without particles and velocities of the 56 particles: $n=0.7$ at 
$t=42.46$ (top), $n=0.5$ at $t=31.78$ (middle), and $n=0.4$ at $t=27.41$ (bottom). }\label{fig.7}	
\end{figure}

\begin{figure}[th!]
\begin{center}
\leavevmode
\includegraphics[width=3.6in]{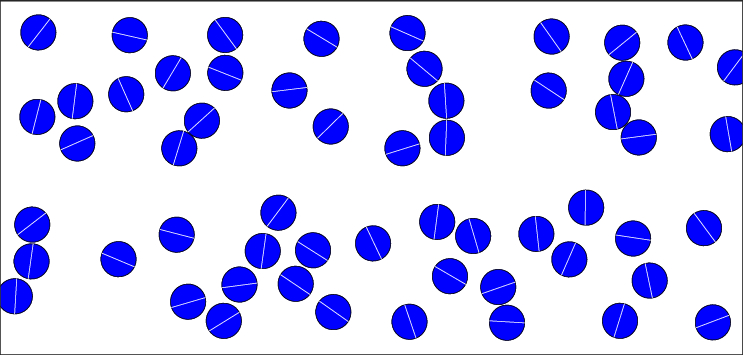}\\
\vskip 2pt
\includegraphics[width=3.6in]{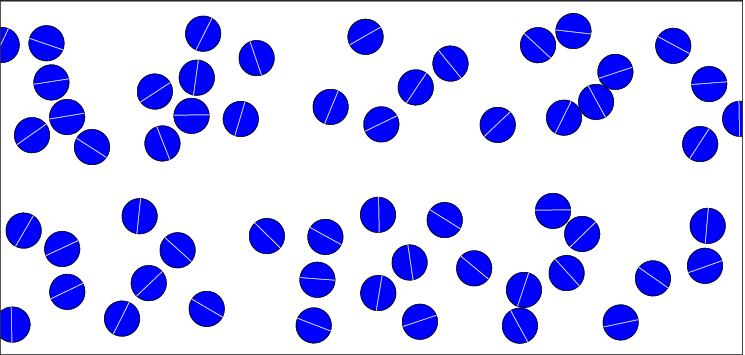}\\
\vskip 2pt
\includegraphics[width=3.6in]{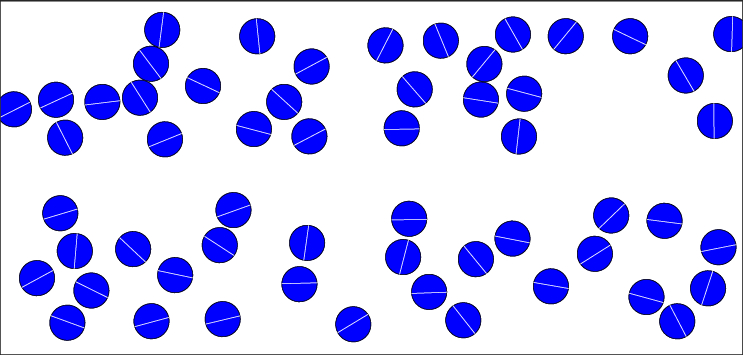}	
\end{center}
\caption{Positions of the 56 neutrally buoyant particles in the pressure driven flow of a shear 
thinning fluid: $n=0.7$ at $t=42.46$ (top), $n=0.5$ at $t=31.78$ (middle), and $n=0.4$ at 
$t=27.41$ (bottom).   }\label{fig.8}	
\end{figure}

\subsection{On the migration of 56 disks in the Poiseuille flow of a shear thinning fluid}

The second validation case we consider concerns the motion and cross stream migration of 56  
neutrally buoyant identical circular cylinders in the pressure driven Poiseuille flow of a 
on-Newtonian shear thinning fluid, which was a test problem studied numerically in \cite{huang2000}. 
The proposed numerical scheme has been validated by comparing our results with those 
reported in  \cite{huang2000}. The computational domain is $\Omega=(0, 21)$ $\times$ $(0, 10)$.
The initial flow velocity and particle velocities are ${\bf 0}$. The pressure drop is $F_1=-2$ 
so that the maximum horizontal speed is 25 when there is no particle for the Poiseuille flow 
of a Newtonian fluid with the viscosity $\eta_0=1$. For non-Newtonian shear thinning fluids 
modeled by (\ref{eqn:2.6a}), the viscosity at infinity shear rate is $\eta_{\infty}=0.1 \eta_0$ 
and the relaxation time is $\lambda_3=1$.  The values of the power index $n$ are 0.4, 0.5, and 0.7.
The fluid and particle densities  are 1. The disk diameter  $d$ is 1  and $\bu_p(t)$ is 
${\bf 0}$ for each disk. The mesh size for the velocity field is $h_v=1/16$, while the mesh 
size for the pressure is $h_p=2 h_v$. The time step is $\triangle t=0.001$.

For the case of a Newtonian fluid without shear thinning ($n=1$), the migration of the disks 
is shown in Figure  \ref{fig.6}. There are no disks at the center line and the disks tend 
to accumulate a distance of 0.6 from the center line of the channel, which is  known as the 
Segre-Silberberg effect (see \cite{segre1961} and \cite{segre1962}). In general, lubrication 
forces move the particles away from the channel walls while the curvature of the velocity profile 
moves the particles away from the centerline of the channel. For a small neutrally buoyant ball 
in a plane Poiseuille flow via the matched asymptotic expansion methods, it was found that the 
of the fluid without particles and the velocities of the particles at $t=100$ are shown in 
Figure \ref{fig.6}. At $t=100$, the maximum particle velocity is 14.0864, while
the maximum fluid velocity without particles is 25 so the slip
velocity is 10.9136 near the centerline. In \cite{huang2000}  the maximum particle velocity
is 15 near the centerline and the maximum slip velocity is 10. 
Both results are in good agreement.  

The velocity profile with shear thinning is quite different from the one at $n=1$ as shown in 
Figure \ref{fig.7}.  The effects of shear thinning can be increased by decreasing the power index $n$.
The maximum particle velocities are 23.0664, 29.9960, and 30.1694, while the maximum fluid velocities 
are 47.3956, 84.5356, and 111.9036 for $n=0.7$, 0.5, and 0.4, respectively. Hence we obtain the slip  
velocities in Table 2 by following the approach given in  \cite{huang2000}. Their slip velocities are 23, 
54, and 81.4  for $n=0.7$, 0.5, and 0.4, respectively, implying our numerical results are in a good agreement
with those in  \cite{huang2000}. Due to the Segre-Silberberg effect, the disks separate into two groups
(see Figures \ref{fig.7} and \ref{fig.8}), the top group and the bottom one according to the particle
positions shown in Figure \ref{fig.8}. Following these particle separations, 
there are three particle  free zones, the one in the middle (called middle zone), the one next to the top 
wall (called the top boundary zone), and the one next to the bottom wall (the bottom boundary zone), which 
can be identified easily from the particle velocity plots in Figure \ref{fig.7}. The sizes of these 
particle free zones are called the gaps and shown in Table 2, in which the average boundary gap
is the mean of the two boundary gaps. 
Due to the shear thinning effect, the disk accumulation is enhanced so that the middle gap of particle
free zone increases and the average boundary gap decreases, when decreasing the value of the power index $n$.
Our results are consistent with those obtained in  \cite{huang2000} qualitatively.
\begin{center}
\hskip 20pt \vbox{\tabskip=0pt \offinterlineskip
\def\tablerule{\noalign{\hrule}}
\halign to 303pt{\vrule height 12pt depth 10pt width0pt#& \vrule#\tabskip=0.35em plus0em&
#& \vrule#&  #& \vrule#&  #& \vrule#&  #& \vrule#&  #& \vrule#&  #& \vrule#&   #& \vrule#& #& \vrule#\tabskip=0pt\cr\tablerule
%each {\hfil#& \vrule#&} align one vertical line
&& \hfil $n$ \hfil && \hfil Slip velocity \hfil &&\hfil   Average boundary gap \hfil   &&\hfil   Middle gap  \hfil &\cr\tablerule
&& \hfil   1    \hfil  && \hfil  10.9136  \hfil && \hfil  0.95167 \hfil  &&  \hfil  1.60403  \hfil   & \cr\tablerule
&& \hfil   0.7  \hfil  && \hfil  24.3292  \hfil && \hfil  0.85906 \hfil  &&  \hfil  1.67752  \hfil   & \cr\tablerule
&& \hfil   0.5  \hfil  && \hfil  54.5396  \hfil && \hfil  0.83388 \hfil  &&  \hfil  1.80585  \hfil   & \cr\tablerule
&& \hfil   0.4  \hfil  && \hfil  81.7342  \hfil && \hfil  0.83313 \hfil  &&  \hfil  2.00206  \hfil   & \cr\tablerule
}}
\vskip 1ex
\begin{minipage}{6in}
Table 2.  The slip velocity,  average of gap sizes next to two walls, and  gap size in the
center of the channel for the power indices $n=1/3$, 1/2, 2/3, and 1.
\end{minipage}
\end{center}

\subsection{On the motion of a self-propelled swimmer}

\begin{figure}[!tp]
\begin{center}
\hskip -50pt \includegraphics[width=0.75true in]{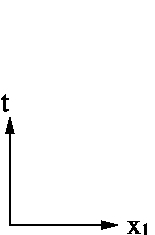}\hskip 50pt 
\includegraphics[height=2.3in]{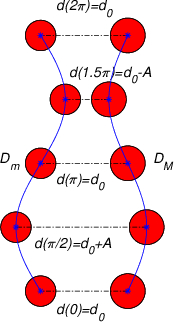}
\end{center}
\caption{An example of the relative position of the two disks: The reciprocal motion with respect to the 
swimmer mass center at different time over one period is $d(t)=d_0+A \sin(t)$.  Disk centers of mass (blue *) are indicated.  }  \label{fig.9}	
\end{figure}

Even Purcell's scallop theorem  \cite{purcell1977} rules out reciprocal motion (i.e., strokes 
with time-reversal symmetry) for effective locomotion in the absence of inertia ($Re=0$) in 
Newtonian fluids, the motion of a simple reciprocal model swimmer (an asymmetric dumbbell) 
in a Newtonian fluid at intermediate Reynolds numbers has been studied extensively in
\cite{dombrowski2019} and \cite{dombrowski2020}. In this section, we have considered  
the case of a neutrally buoyant  self-propelled swimmer freely  moving in a non-Newtonian 
shear thinning fluid.  Such swimmer is formed by two disks of different sizes connecting by 
a massless spring at their mass centers (see Figure \ref{fig.9}). Hence it is an asymmetric
long body, but symmetric with respect to the line segment connecting  two disk mass centers. 
The reciprocal motion of two disks with respect to the swimmer mass center is prescribed 
(see Figure \ref{fig.9} and equation (\ref{eqn:3.28})). Due to the asymmetric disk motion 
with respect to the swimmer mass center, swimmer can  move in either directions parallel to the string 
connecting the two disk mass centers.  It is interesting to find out how the shear thinning impacts 
the swimmer motion. Following the units used in  \cite{dombrowski2019},   we assume   in this 
section  all dimensional quantities are in the physical MKS units.

\begin{figure} [!t]
\begin{center}
\includegraphics[height=2.4in]{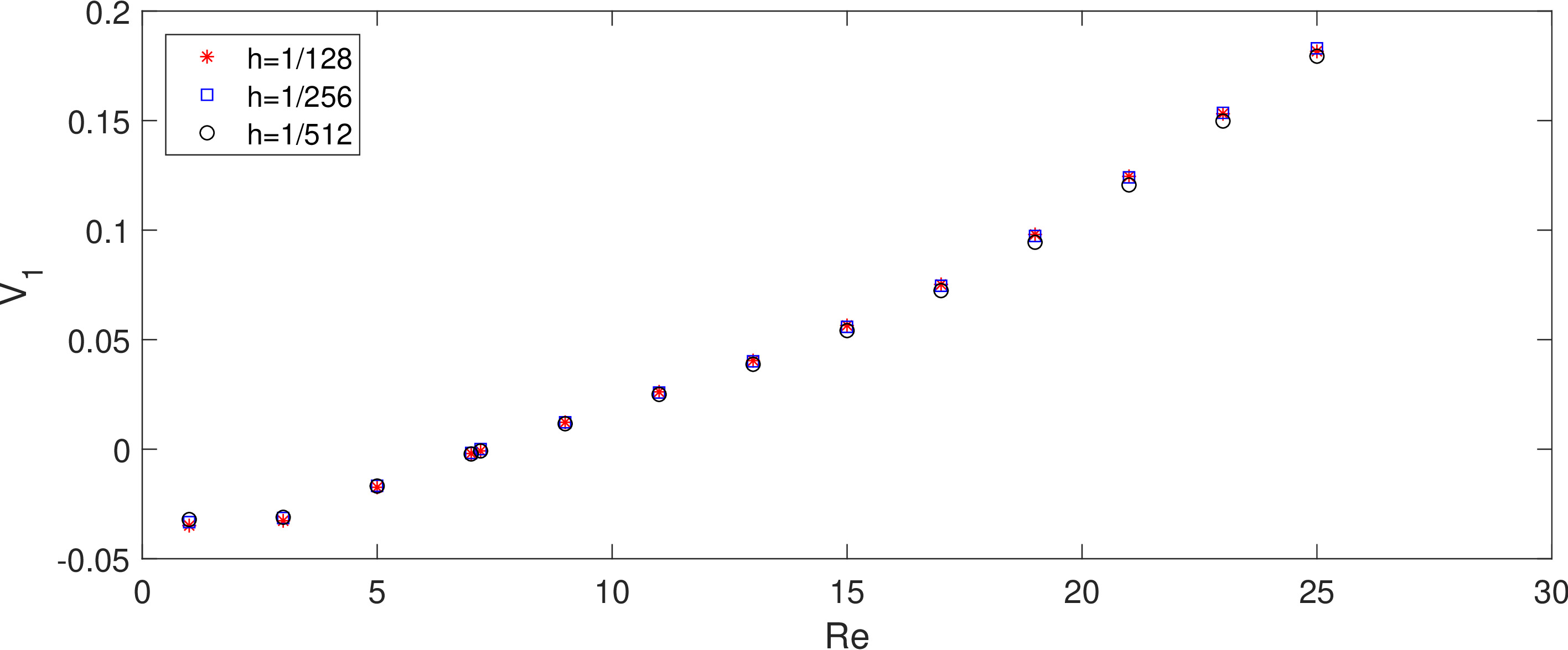}\\
\includegraphics[height=2.4in]{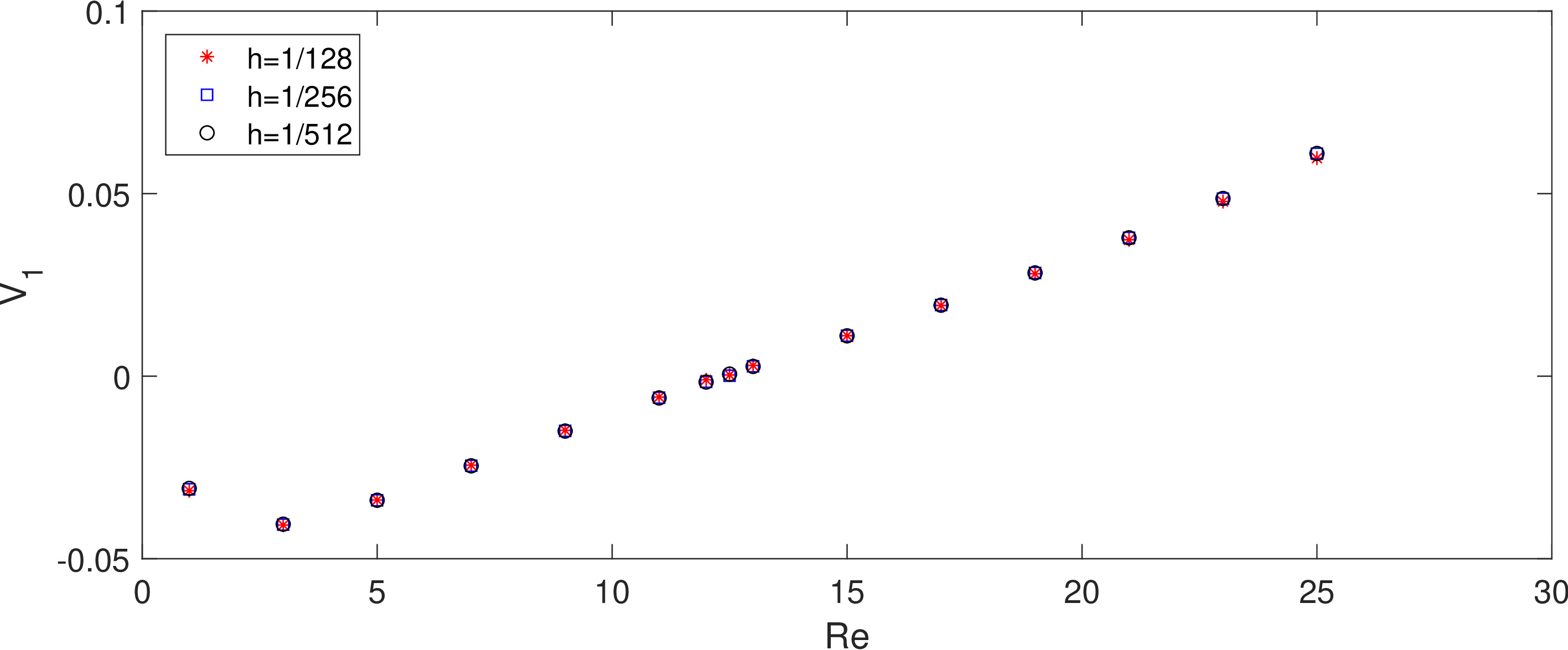}
\end{center}
\caption{Average horizontal velocity $V_1$ of a swimmer versus $Re$ for the power index $n=1$
	(top) and 0.8 (bottom).}  \label{fig.11}	
\end{figure}

\begin{figure}[!t]
\begin{center}
\includegraphics[height=1.4in]{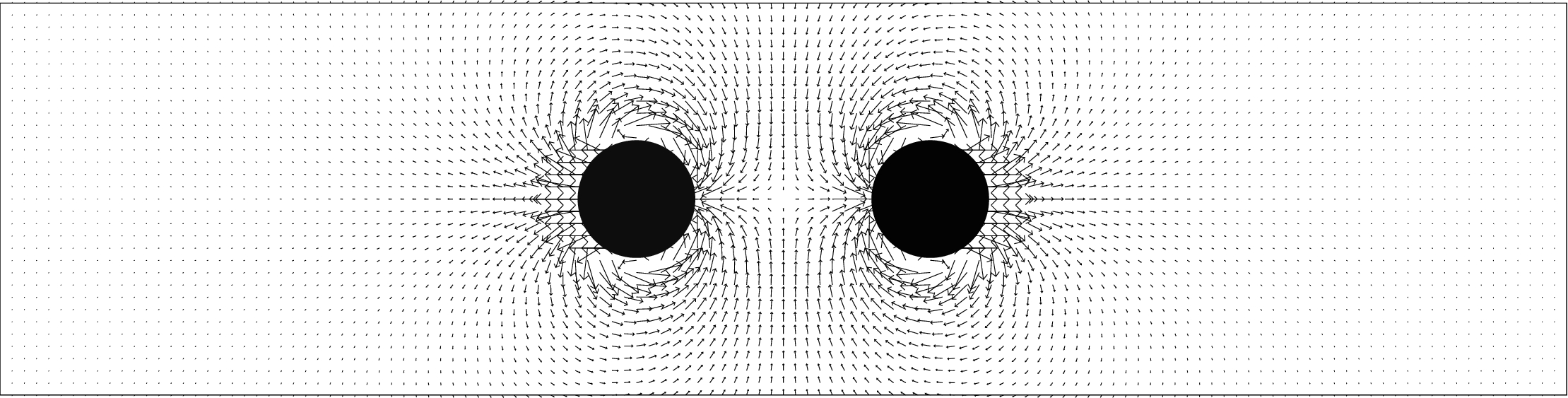}\\
\vskip 2pt                    
\includegraphics[height=1.41in]{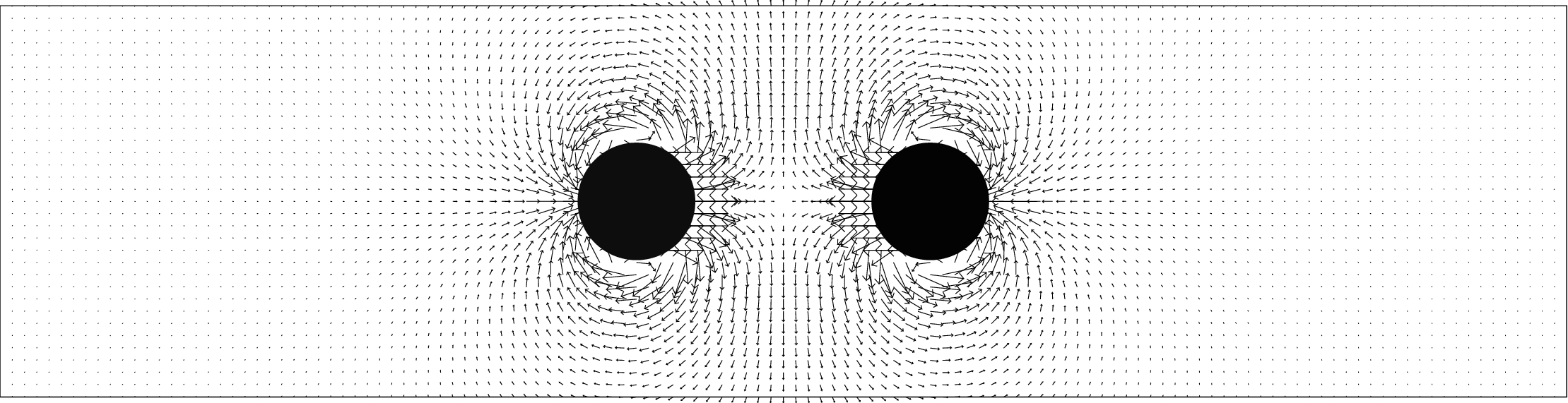}
\end{center}
\caption{Velocity field next to the swimmer of two same size disks and swimmer position at 
$t=8$ (top) and 8.05 (bottom) for $Re=15$ and $n=1$.}  \label{fig.14}	
\end{figure}
\begin{figure}[!t]
\begin{center}
\hskip -10pt \includegraphics[height=2.5in]{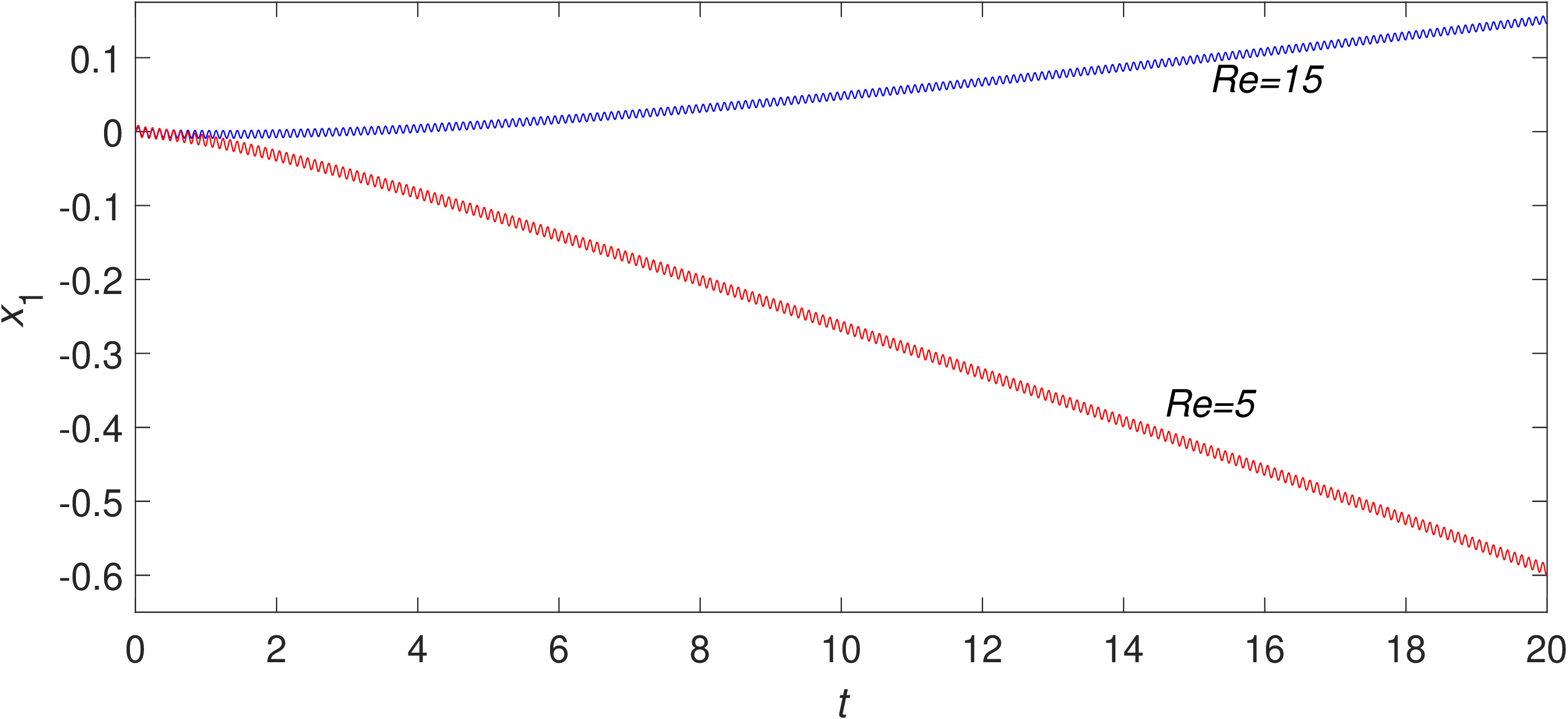}
\end{center}
\caption{Histories of the horizontal $x_1$-position   of the swimmer mass center for $Re=5$ 
and 15 and $n=1$ where the swimmer is formed by two different size disks (see Figure 
\ref{fig.16}).}  \label{fig.15}	
\end{figure}
The computational domain is $\Omega=(-4, 4)$ $\times$ $(-4, 4)$. The flow velocity is ${\bf 0}$ 
initially and zero at the boundary of domain $\Omega$. 
The densities of fluid and swimmer are $10^3$, which is the water density.
The initial position of the swimmer mass center is at (0,0). The distance between the two disk 
centers is $d(t)=d_0+A\sin(\omega t)$ as  in \cite{dombrowski2019} where the equilibrium distance 
between the two disk centers is $d_0=10 r_0$, the amplitude is $A=r_0$, and $r_0=0.15$.  
The angular frequency is $\omega=20 \pi$ (so the frequency is 10). The massless spring connecting 
the two disks is located in the direction parallel to the $x_1$ direction. The radius of the 
larger disk $D_M$ is $R_M=2r_0=0.3$ and that of the smaller disk $D_m$ is $R_m=r_0=0.15$. 
The amplitude $A$ is $A=A_M+A_m$ where the component for the larger disk is 
$A_M=\dfrac{A R_m^2}{R_M^2+R_m^2}$ and the one for the smaller disk is  
$A_m=\dfrac{A R_M^2}{R_M^2+R_m^2}$.  The Reynolds number is defined by 
$Re=A_m R_m \omega/\nu$ (as in \cite{dombrowski2019}) where $\nu$ is the kinetic viscosity of 
the fluid at zero shear rate. We vary the value of $Re$ to obtain the fluid viscosity at zero 
shear rate. We have used structured triangular meshes 
in all simulations.  The mesh size for the velocity field is $h_v=1/128$, while the mesh size 
for pressure is $h_p=2 h_v$. The time step is $\triangle t=0.001$. Concerning the shear 
thinning, the ratio of viscosity  at infinity shear rate and  the one at zero shear 
rate is 0.1 and the relaxation time is $\lambda_3=1$.  The values 
of the power index $n$ are 0.5, 0.6, 0.7, 0.8, and 0.9. 
\begin{figure}[!t]
\begin{center}
\includegraphics[height=1.3in]{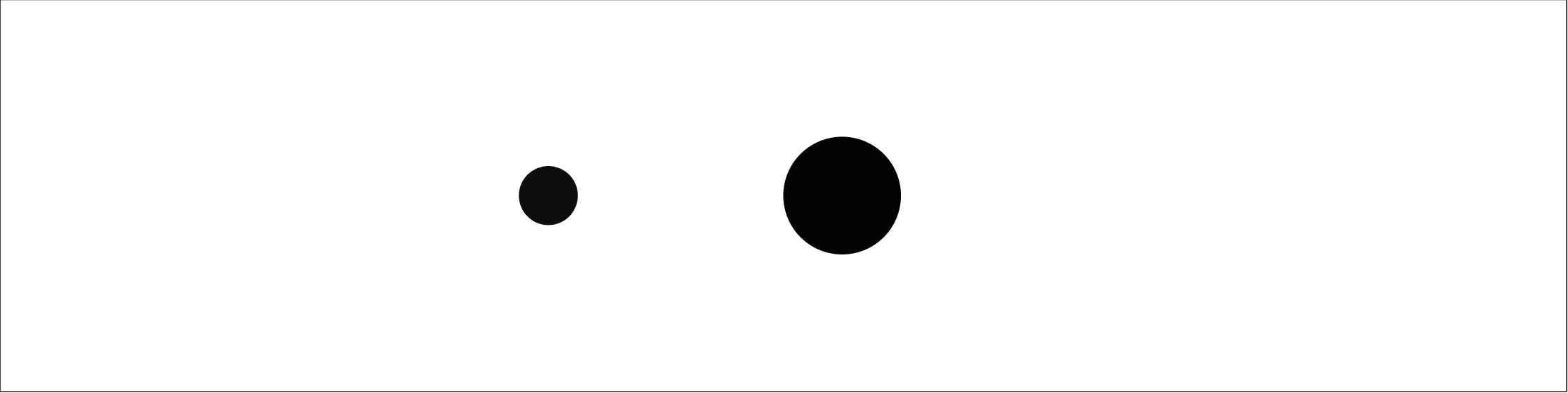} \\
\vskip 2pt
\includegraphics[height=1.3in]{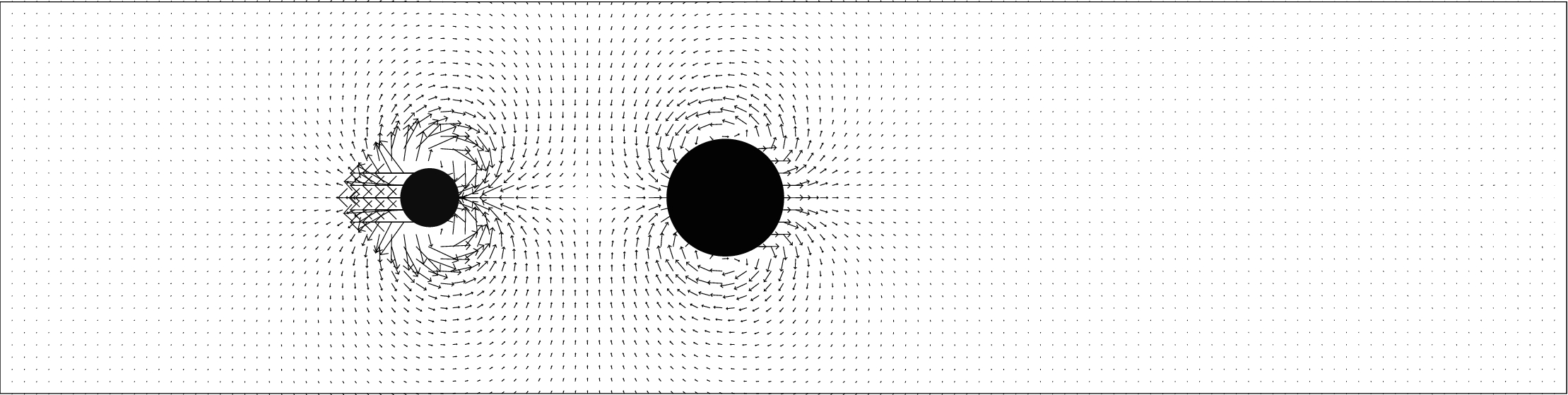} \\
\vskip 2pt
\includegraphics[height=1.3in]{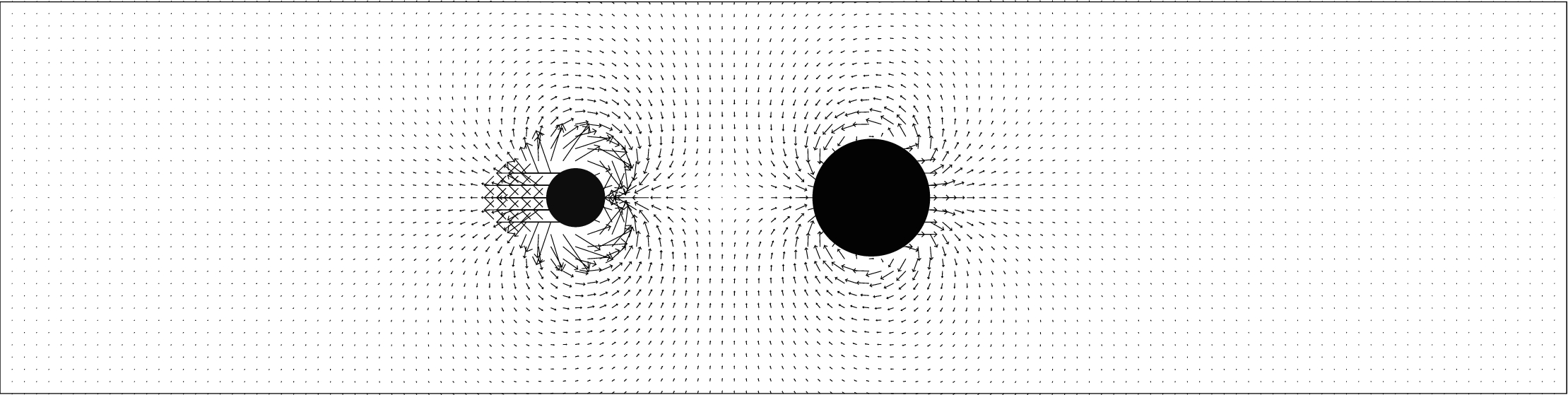}
\end{center}
\caption{Swimmer position at $t=0$ (top), velocity field next to the swimmer and  swimmer position 
at $t=20$ for $Re=5$ (middle) and 15 (bottom) and $n=1$.}  \label{fig.16}	
\end{figure}

In  this section, we have mainly studied the effect of the power index $n$ and of the Reynolds 
number on the moving direction and speed of the swimmer. In order to focus on the swimmer moving 
direction, the swimmer is only allowed to freely move in the $x_1$-direction. 
To accommodate this restricted motion and the Dirichlet boundary condition, 
we have to modify the spaces defined previously  as follows
\begin{eqnarray*}
W_{0} &=& \{\bv|\bv \in (H^1(\Omega))^2, \ \bv = {\bf 0} \ 	\text{\it on the boundary of $\Omega$.}   \},\\
\Lambda_0(t) &=& \{\bmu| \bmu \in (H^1(B(t)))^2, <\bmu,{\bf e}_1>_{B(t)}=0 \},\\
W_{0,h}&=&\{\bv_h | \bv_h \in (C^0(\overline{\Omega}))^2, \  \bv_h|_T \in P_1\times P_1,  \forall T\in \ct_h, \\
 && \bv_h = {\bf 0} \  \text{\it on the boundary} \ \text{\it  of $\Omega$}  \},  \\
%\end{eqnarray*}
%\begin{equation*}
L^2_{0,h}&=&\{q_h|q_h\in  C^0(\overline{\Omega}), \ q_h|_T\in P_1,
\ \forall T\in \ct_{2h},  \ \int_{\Omega} q_h \, d\x=0 \},\\
%\end{equation*}
%\begin{equation*}
\Lambda_{0,h}(t) &=&\{\mu_h| \mu_h \in \Lambda_h(t), \ <\mu_h,{\bf e}_1>_{B_h(t)}=0 \}.
%\end{equation*}
\end{eqnarray*}
Following the above given restricted motion, the reciprocal motion with respect to the swimmer mass center is
\begin{equation}
\bu_p(t)=\begin{cases}
(A_M\omega \cos(\omega t),0)^T,  \ \forall \x \in D_M,\\
(-A_m\omega \cos(\omega t),0)^T,  \ \forall \x \in D_m.\\
\end{cases}\label{eqn:3.28}
\end{equation}
\begin{figure}[]
\begin{center}
\includegraphics[height=2.4in]{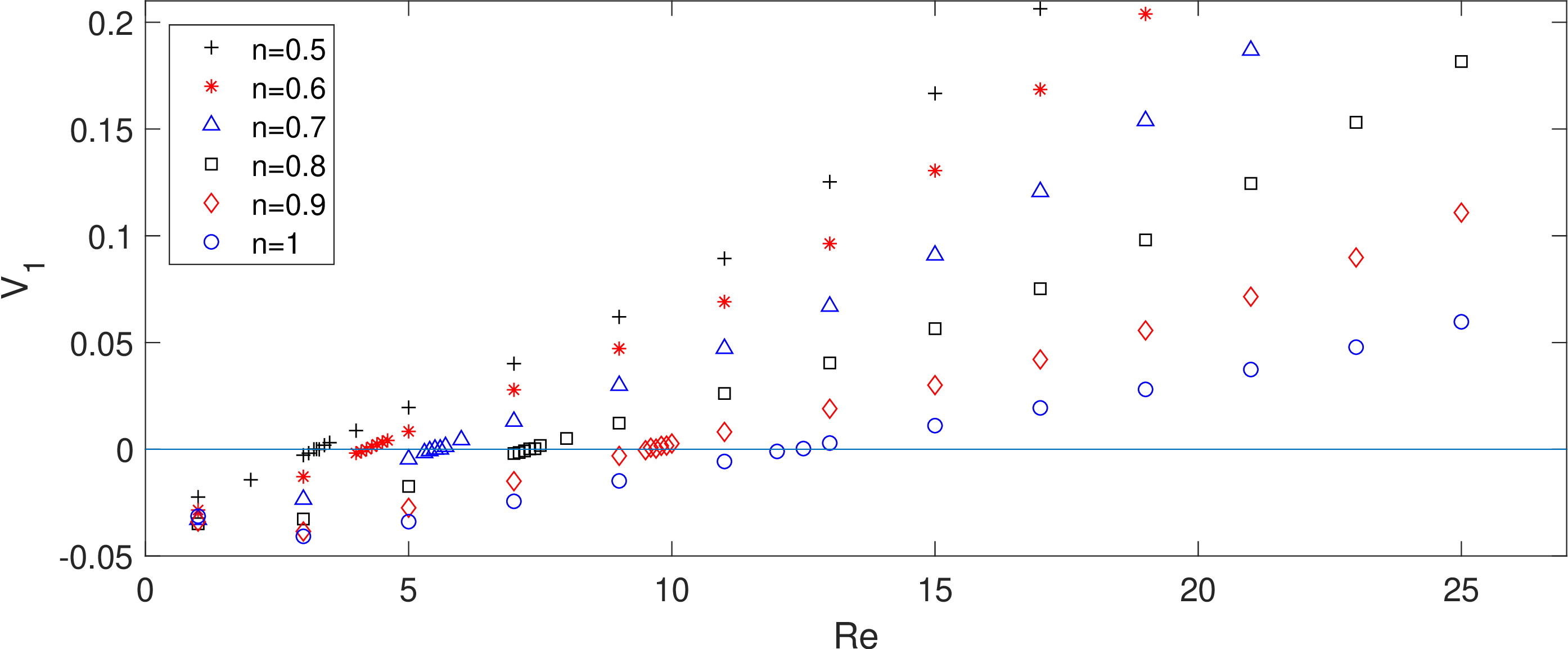}
\end{center}
\caption{Average horizontal speed $V_1$ of a swimmer versus $Re$ for the power index $n=0.5$,   0.6,
0.7, 0.8,  0.9, and 1.}  \label{fig.17}	
\end{figure}
\begin{figure}[!t]
\begin{center}
\includegraphics[height=1.3in]{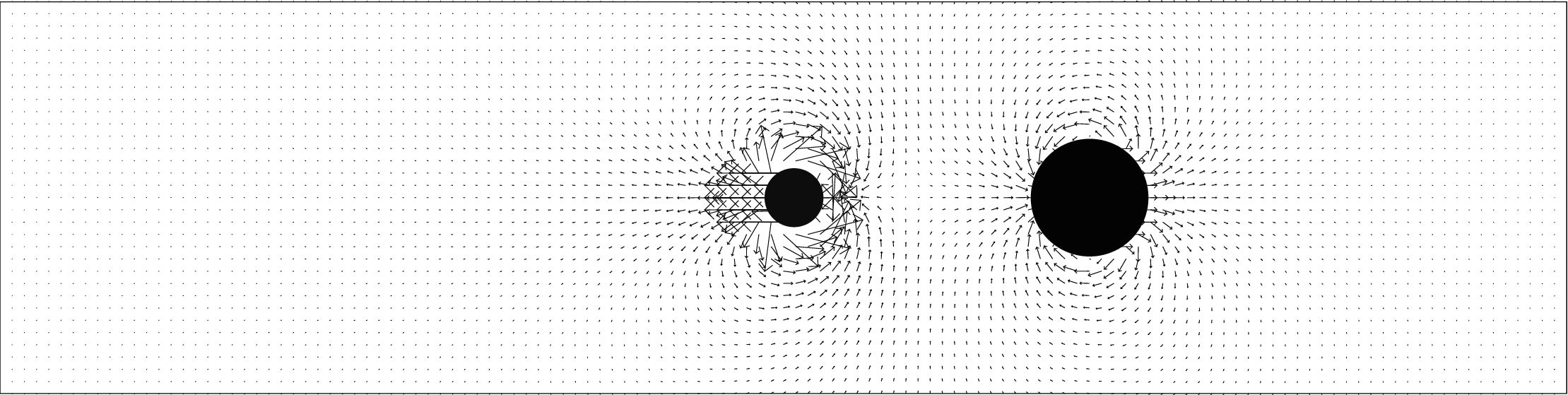}\\
\vskip 2pt
\includegraphics[height=1.3in]{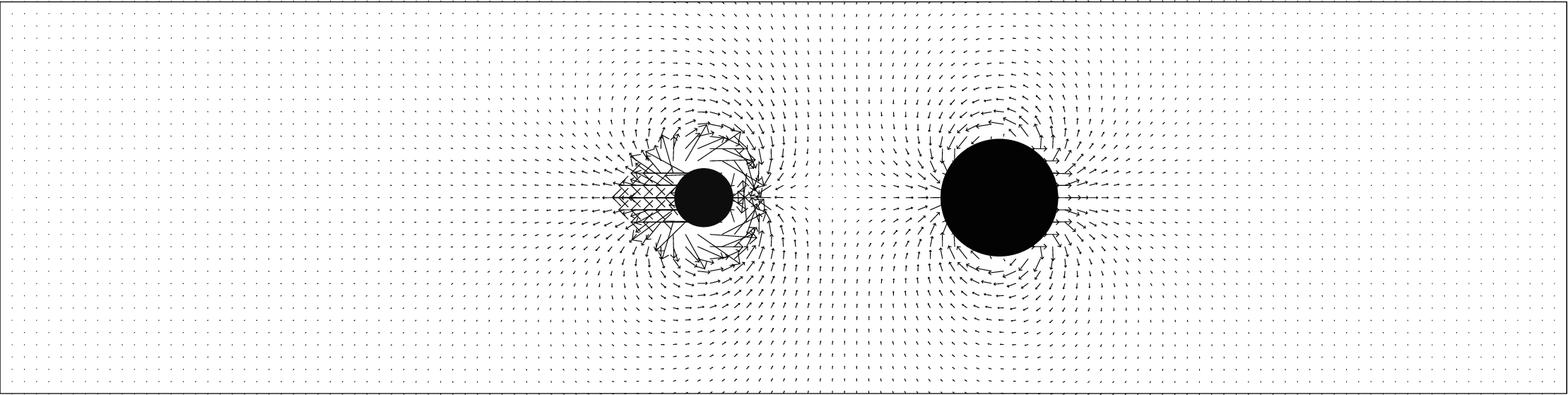}\\
\vskip 2pt
\includegraphics[height=1.3in]{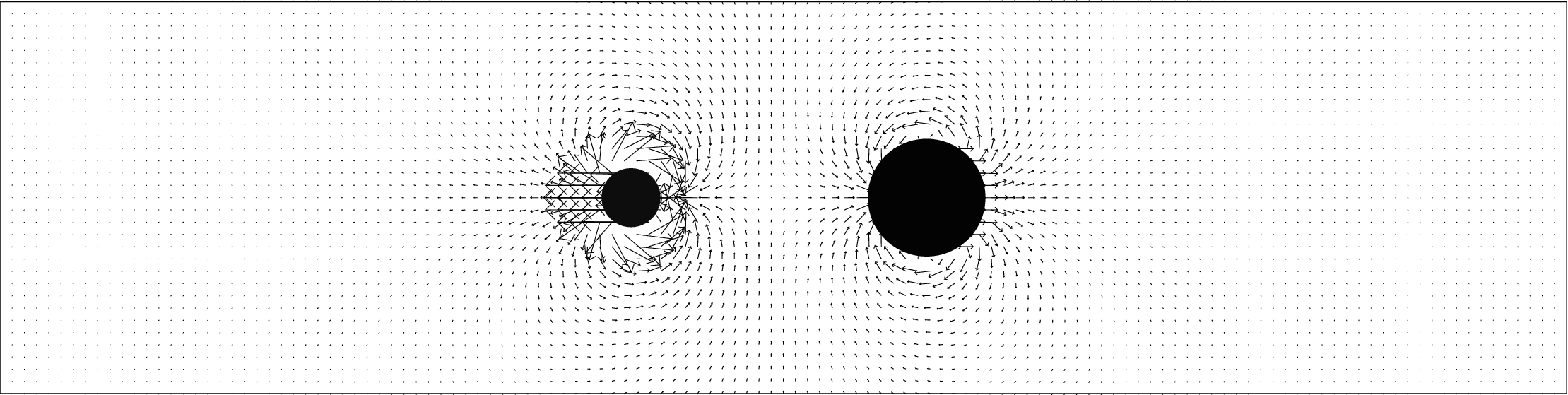}		
\end{center}
\caption{Velocity field next to the swimmer and swimmer position at $t=20$ for $Re=15$ 
and $n=0.7$ (top), $n=0.8$ (middle), and $n=0.9$ (bottom).}  \label{fig.18}	
\end{figure}
Applying algorithm  (\ref{eqn:3.12})-(\ref{eqn:3.27}) (with the subproblem  (\ref{eqn:3.16a}) 
instead of problem (\ref{eqn:3.16}) for shear thinning fluids) and above discrete spaces, we have 
studied numerically the motion of two disk swimmer in the $x_1$ direction in Newtonian and 
non-Newtonian fluids. We first show the average velocities of swimmer for different values 
of $Re$ and two power index values, $n=1$ and 0.8, obtained by the velocity mesh sizes 
$h=1/128$, 1/256, and 1/512 and time step $\triangle t=0.001$. Each average value was 
computed from the last twenty periods of horizontal velocity for  $0\le t \le 20$.
All those average velocities for each Reynolds number are in a good agreement as 
shown in Figure \ref{fig.11}.

Let us  consider a swimmer of two same size disks, i.e., $R_m=R_M=2r_0=0.3$, 
suspended in a Newtonian fluid ($n=1$). Due to the symmetry of reciprocal motion of the 
two same size disks (see Figure \ref{fig.14}), the swimmer mass center does not move as expected, 
at least for $Re=1$, 3, 5, 7, 9, 11, 13, and 15, the cases  we have considered. But for 
the case of two different size disks in a Newtonian fluid with $R_M=2 r_0$ and $R_m=r_0$ 
for $r_0=0.15$, the swimmer can move in the positive or negative $x_1$-direction up to the 
Reynolds number.  As shown in Figure \ref{fig.15}, the swimmer mass center oscillates 
with period 0.1 but moves in the positive (resp., negative) $x_1$-direction for $Re=15$ 
(resp., $Re=5$). There is a critical Reynolds number so that the swimmer moves to the left 
(resp., right) for  $Re < Re_c$  (resp., $Re > Re_c$) if the larger disk is located on the 
right side (see the plots in Figures \ref{fig.16} and \ref{fig.17}).  The critical Reynolds 
number is about $Re_c=12.5$ for the swimmer shown in  Figure \ref{fig.16}. 

But when swimming in a non-Newtonian shear thinning fluid, 
the average moving speed ($V_1$) per period in the positive-$x_1$ 
direction increases when decreasing the value of the power index $n$ as shown in Figures 
\ref{fig.17} for the power index values, $n=0.5$, 0.6, 0.7, 0.8, 0.9, and 1. Interestingly the 
average swimming velocity is also slow down when the swimmer moves in the minus-$x_1$ direction
when  decreasing the value of the power index $n$. Thus the swimmer position is further to the right
when comparing its position at $t=20$ as in Figure \ref{fig.18}. Our computational results show that
the shear thinning does have a strong effect on the motion of a swimmer made of two different size 
disks and the critical Reynolds number decreases for a non-Newtonian shear thinning fluid as the 
power index value decreases.

\section{Conclusions}
In this article, a distributed Lagrange multiplier/fictitious domain method has been developed 
for simulating a non-symmetric (two-disk) particle moving freely in non-Newtonian shear 
thinning fluids. The Carreau-Bird model is adapted for the fluid shear thinning property. Our 
numerical methods have been validated by comparing our computed solutions with the recently 
published exact solutions for Poiseuille flows  of non-Newtonian fluids in two dimensions 
(in ref. \cite{griffiths2020}). 
Accurate numerical solutions have been obtained and the $L^2$-errors of the velocity field show 
the expected 2nd order for $P_1$ finite elements. For many neutrally buoyant particle cross stream 
migration in a Poiseuille flow of non-Newtonian shear thinning fluids, the particles migrate away
from the wall and center of the channel. The size  of the boundary gap decreases and that of the 
middle gap increases when reducing the power index value. These results are consistent qualitatively
with those reported in \cite{huang2000}. Concerning the two-disk swimmer,
the effect of shear thinning does let the swimmer move faster when moving    in the positive
$x_1$-direction, i.e., from the smaller to larger disk,  and the critical Reynolds number (for changing the moving direction from the left 
to the right) decreases for smaller value of the power index.

\vskip 4ex
\noindent{\large\bf Acknowledgments}
\vskip 2ex

We acknowledge the helpful comments and suggestions of  Howard H. Hu (University of Pennsylvania).

\end{document}